\documentclass[11pt,onecolumn,oneside,aps,pra,tightenlines,notitlepage,nofootinbib,superscriptaddress,floatfix]{revtex4-2}

\usepackage{amssymb}
\usepackage{amsmath}
\usepackage{amsbsy}
\usepackage{amsfonts}
\usepackage{mathrsfs}
\usepackage{amsthm}
\usepackage{mathtools}

\usepackage{graphicx}

\usepackage{tikz}
\usepackage{tikz-cd}
\usetikzlibrary{quantikz2}

\usepackage{xcolor}
\definecolor{darkred}  {rgb}{0.5,0,0}
\definecolor{darkblue} {rgb}{0,0,0.5}
\definecolor{darkgreen}{rgb}{0,0.5,0}

\usepackage{hyperref}
\hypersetup{
  breaklinks = true,
  colorlinks = true,
  urlcolor = blue,
  linkcolor = darkblue,
  citecolor = darkgreen,
  filecolor = darkred
}

\theoremstyle{plain}

\theoremstyle{definition}
\newtheorem{definition}{Definition}

\renewcommand{\leq}{\leqslant}

\newcommand{\Int}{\mathbb{Z}}

\DeclareMathOperator{\Tr}{Tr}

\DeclarePairedDelimiter{\abs}{\lvert}{\rvert}

\DeclarePairedDelimiterX{\innerp}[2]{\langle}{\rangle}{#1, #2}

\DeclarePairedDelimiter{\ket}{\lvert}{\rangle}
\DeclarePairedDelimiterX{\braket}[2]{\langle}{\rangle}{#1 \delimsize\vert #2}
\DeclarePairedDelimiterX{\ketbra}[2]{\lvert}{\rvert}{#1 \delimsize\rangle\!\delimsize\langle #2}
\DeclarePairedDelimiterX{\proj}[1]{\lvert}{\rvert}{#1 \delimsize\rangle\!\delimsize\langle #1}

\newcommand{\NOT}{\mathrm{NOT}}
\newcommand{\CNOT}{C\NOT}

\newcommand{\BigO}{\mathcal{O}}

\newcommand{\Hil}{\mathcal{H}}

\newcommand{\TrClass}{\mathfrak{T}}

\newcommand{\Id}{\mathrm{Id}}
\newcommand{\Deph}{\mathcal{D}}

\begin{document}

\title{A streamlined demonstration that stabilizer circuits simulation reduces to Boolean linear algebra}

\author{Vsevolod I. Yashin}
\email{yashin.vi@mi-ras.ru}
\affiliation{Steklov Mathematical Institute of Russian Academy of Sciences, Moscow 119991, Russia}
\affiliation{Russian Quantum Center, Skolkovo, Moscow 143025, Russia}
\affiliation{Moscow Institute of Physics and Technology, Dolgoprudny 141700, Russia}

\date{\today}

\begin{abstract}
  Gottesman-Knill theorem states that computations on stabilizer circuits can be simulated on a classical computer, conventional simulation algorithms extensively use linear algebra over bit strings. For instance, given a non-adaptive stabilizer circuit, the problem of computing the probability of a given outcome (strong simulation) is known to be log-space reducible to solving the system of linear equations over Boolean variables, which is commonly done by Gaussian elimination. This note aims to make the connection between stabilizer circuits and Boolean linear algebra even more explicit. To do this, we extend the stabilizer tableau formalism to include stabilizer tableau descriptions of arbitrary stabilizer operations (Clifford channels). Finding the tableau corresponding to the composition of two channels becomes a linear algebra problem. Any stabilizer circuit rewrites to a diagram with stabilizer tableaux on vertices, contracting an edge means to take the composition of channels, to compute the result of the circuit means to fully contract the diagram. Thus, simulating stabilizer circuits reduces to a sequence of Gaussian eliminations. This approach gives a new perspective on explaining the work of stabilizer tableau methods (reproducing the asymptotics) and creates opportunity for exploring various tensor-contraction techniques in stabilizer simulation.
\end{abstract}

\maketitle

\section{Introduction} \label{sec:introduction}

Studying stabilizer circuits is an important topic in quantum computation and quantum error correction. Stabilizer operations were introduced by Gottesman to characterize a class of operations that are natural to implement fault-tolerantly in error-correcting codes \cite{Gottesman_1997, Gottesman_1998}. Fault-tolerant implementation of non-stabilizer operations requires additional techniques (such as magic state distillation \cite{Bravyi_2005, Gidney_2024}) and are considered costly. At the same time, it was found that stabilizer circuits can be efficiently simulated on classical computers (Gottesman-Knill theorem \cite{Nielsen_2010}), which means that they cannot demonstrate computational quantum advantage, but which also ensures that one can succesfully study the propagation of errors in circuits and create efficient syndrome decoding algorithms for error correcting codes.

Nice properties of stabilizer circuits create an incentive for computer software develompent. Indeed, today there are quite a few projects working in the area of stabilizer simulation. Various approaches include: stabilizer tableau methods \cite{Gottesman_1997, Gottesman_1998, Aaronson_2004, Gidney_2021}, graph state representations \cite{Anders_2006, Rijlaarsdam_2020, Hu_2022, Khesin_2025}, quadratic form expansions \cite{Van_den_Nest_2010, Bravyi_2016, Dehaene_2003, Beaudrap_2022}, quasiprobability representations and hidden variable models \cite{Gross_2006, Catani_2017, Park_2023}. One can distinguish between two problems of simulation: weak and strong \cite{Van_den_Nest_2010}. Weak simulation is the problem of sampling from the outcome distribution of a circuit, strong simulation is the problem of computing the probabilitiy of a given outcome. Usually, weak simulation is efficienlty reducible to strong simulation \cite{Bravyi_2022}, on the other hand the outcome probability can be estimated from statistics over many samples. In \cite{Aaronson_2004} it was shown that the problem of strong simulation of stabilizer circuits without classical control is $\oplus\mathtt{L}$-complete and log-space reducible to the problem of solving a linear system of equations with Boolean variables. Such problems are exactly those efficiently solvable on affine Boolean circuits. Therefore, all known simulation algorithms extensively use linear algebra over Boolean variables. Stabilizer circuits are computationally equivalent to usual Boolean circuits, meaning that one can only simulate them weakly. Thorough investigation on various classes of simulatble circuits was done in \cite{Jozsa_2013, Koh_2017, Bouland_2018}.

Even though stabilizer circuits do not exhibit computational advantages of general quantum circuits, most protocols of quantum communications are in fact stabilizer, therefore it is fruitful to study the general theory of stabilizer  operations. There is a lot of great research on characterizing entanglement of stabilizer states and graph states in particular \cite{Fattal_2004, Looi_2011, Hein_2004}. Also, more recently appeared a number of works on characterising the general properties of stabilizer operations \cite{Heimendahl_2022, Kliuchnikov_2023, Yashin_2025}.

In this work we aim to unify the principles of stabilizer simulation and to provide a better pedagogical explanation for them. Firstly, we develop the notion of stabilizer tableau to include describing arbitrary stabilizer operations (Clifford channels). Basically, we propose using the modification of the Choi state stabilizer tableau. We comment on the properties of such tableau formalism and find that the problem of composing two Clifford channels reduces to the problem of finding a basis in the intersection of two vector spaces. Then, we show that a circuit can be understood as a diagram with stabilizer tableux of vertices and connecting wires on edges, contracting an edge corresponds to composition of channels. Simulaiton of the circuit can be done by fully contracting the diagram. The strategies for diagram contraction can vary, simple strategies reproduce known simulation algorithms.

The work is divided into two parts. In Section~\ref{sec:extending} we remind basic notions and extend the stabilizer tableau formalism, in Section~\ref{sec:explaining} we study the examples, discuss the problem of stabilizer simulation algorithms and give an outlook.

\section{Extending stabilizer tableau formalism} \label{sec:extending}

In this Section we discuss stabilizer tableau formalism and extend it to include descriptions of arbitrary non-adaptive stabilizer operations. We start by fixing some notation and reminding about the ordinary stabilizer tableau formalism as established in founding papers \cite{Gottesman_1997, Gottesman_1998, Aaronson_2004}: describing properties of Pauli operators and explaining the idea of stabilizer tableau, which represents stabilizer state as Boolean matrix. After that, we discuss the notion of Clifford channels and introduce a way to encode them by modified versions of stabilizer tableaux. An in-depth review about the properties of stabilizer operations as understood by the author can be found in \cite{Yashin_2025}.

\subsection{Pauli operators}

We denote multiqubit systems by uppercase latin letters $A,B,C$, the number of qubits in a system $A$ is denoted $\abs{A}$. We denote $\Hil$ some Hilbert space, $\Hil_A$ a Hilbert space over a system $A$; let $\TrClass(\Hil)$ denote the space of trace-class operators on $\Hil$. A \emph{quantum channel} $\Phi : \TrClass(\Hil_A) \to \TrClass(\Hil_B)$ is a completely positive positive trace-preserving linear map between systems $A$ and $B$, we can also write it as $\Phi_{A\to B}$ for clarity. Given a superoperator $\Phi_{A\to B}$ and the maximally entangled pure state $\Omega_{A A}$ on two copies of a system $A$, the \emph{Choi operator} $\sigma$ of that superoperator is defined as $\sigma_{AB} = \Id_{A}\otimes \Phi_{A\to B}[\Omega_{AA}]$. A superoperator and the corresponding Choi operator are related by Choi-Jamio{\l}kowsky duality \cite{Holevo_2019}.

An $n$-qubit \emph{sign-free Pauli operator} is a tensor product of $n$ single-qubit Pauli matrices $\{I,X,Y,Z\}$, and a \emph{Pauli observable} is a Pauli operator together with possible sign $\pm 1$. The Pauli operators with possible phases $\{\pm 1, \pm i\}$ form a \emph{Pauli group}, Pauli observables are exactly the Hermitian elements of Pauli group. The set of sign-free Pauli matrices $\{I,X,Y,Z\}^{\otimes n}$ constitutes orthonormal basis in the space of matrices with Hilbert-Schmidt inner product $(X,Y) \mapsto \frac{1}{2^n}\Tr[X^\dag Y]$.

We will encode $n$-qubit Pauli observables as $(2n+1)$-bit strings. To a $n$-qubit system we correspond a \emph{phase space} $\Int_2^{2n}$. Phase points are enumerated as $2n$-bit strings $u = (z_1,x_1,\dots,z_n,x_n) \in \Int_2^{2n}$. Given a phase space point $u\in\Int_2^{2n}$ and a Boolean value $c\in\Int_2$, we define the corresponding Pauli observables:
\begin{equation}
  P(u) = i^{-\innerp{z}{x}} Z_1^{z_1}X_1^{x_1}\, \cdots \,Z_n^{z_n}X_n^{x_n}, \qquad P(u|c) = (-1)^c P(u),
\end{equation}
where $\innerp{z}{x} = \sum_{i=1}^n z_i x_i$ is an inner product of two bit strings taken modulo $4$. The operators $P(u)$ are sign-free, the additional bit $c$ in $P(u|c)$ represents a sign. Thus, any Pauli observable is uniquely represented by some bit string $(u|c)\in\Int_2^{2n+1}$. The transpose of a Pauli observable is
\begin{equation}
  P(u|c)^T = (-1)^{\innerp{z}{x}} P(u|c) = P(u|c\oplus\innerp{z}{x}).
\end{equation}
The product of two Pauli observables satisfies Weyl-type commutation relation
\begin{equation}
  P(u|c) \cdot P(u'|c') = i^{\beta(u,u')} P(u\oplus u'|c\oplus c'),
\end{equation}
where the bit strings $(u|c), (u'|c')\in\Int_2^{2n+1}$ are added modulo $2$, and an additional scalar phase is described by a $2$-cocycle \cite{Heinrich_2021}
\begin{equation}
  \beta(u,u') = \innerp{z\oplus z'}{x\oplus x'} - \innerp{z}{x} - \innerp{z'}{x'} - 2\innerp{z'}{x}.
\end{equation}
Note that the summations inside of inner product arguments are taken modulo $2$, while the outer summation is modulo $4$, so the function $\beta$ is \emph{not} a bilinear form, but it is antisymmetric $\beta(u,u') + \beta(u',u) = 0$. Two observables $P(u|c)$ and $P(u'|c')$ commute as
\begin{equation}
  P(u|c) \cdot P(u'|c') = (-1)^{[u,u']}P(u'|c')\cdot P(u|c),
\end{equation}
where $[u,u']$ is the symplectic form on $\Int_2^{2n}$:
\begin{equation}
  [u,u'] = \beta(u,u') \;\mathrm{mod}\,2 = \innerp{z}{x'}\oplus\innerp{z'}{x}.
\end{equation}
In case of commuting observables $[u,u'] = 0$ the $2$-cocycle $\beta(u,u')$ is divisible by $2$, so one can update the bit string of the product as
\begin{equation} \label{eq:product_of_observables}
  P(u|c) \cdot P(u'|c') = P(u\oplus u'|\,c\oplus c'\oplus {\scriptstyle \frac{\beta(u,u')}{2} }).
\end{equation}

\subsection{Stabilizer tableaux}

A \emph{stabilizer group} $\mathcal{S}$ is a commutative group of Pauli observables such that $-I \notin \mathcal{S}$. Any stabilizer group is isomorphic to $\mathbb{Z}_2^r$, where $r$ is a \emph{rank} of $\mathcal{S}$ and $0 \leq r \leq n$. Suppose  $\langle P(u_1|c_1),\dots,P(u_k|c_k)\rangle$ is a finite generating set for some stabilizer group $\mathcal{S}$. This set can be stored in computer memory as a $k\!\times\!(2n+1)$ Boolean matrix $[\mathsf{U}|\mathsf{c}]$ called \emph{stabilizer tableau}:
\begin{equation}
  \left[
  \begin{array}{c|c}
    \mathsf{U} & \mathsf{c}
  \end{array}
  \right]
  =
  \left[
  \begin{array}{c|c}
    u_1    & c_1    \\
    \vdots & \vdots \\
    u_k    & c_k    \\
  \end{array}
  \right]
  =
  \left[
  \begin{array}{ccccc|c}
    z_{11} & x_{11} & \cdots & z_{1n} & x_{1n} & c_1    \\
    \vdots & \vdots & \ddots & \vdots & \vdots & \vdots \\
    z_{k1} & x_{k1} & \cdots & z_{kn} & x_{kn} & c_k    \\
  \end{array}
  \right].
\end{equation}
Each qubit corresponds to two column of the tableau. We allow for degenerate stabilizer tableaux, so we will not require $k$ to equal the rank of $\mathcal{S}$. The requirement $-I\notin \mathcal{S}$ implies that there can be no rows of form $[0\cdots0|1]$. The requirement of Pauli observables to commute means that sign-free part $\mathsf{U}$ of bit matrix is symplectic $\mathsf{U} \mathsf{J} \mathsf{U}^T = 0$, where $\mathsf{J}$ is the matrix of symplectic form
\begin{equation}
  \mathsf{J} = \begin{bmatrix} 0 & 1 \\ 1 & 0 \end{bmatrix} \otimes \mathsf{I}_n
  =
  \begin{bmatrix}
    0 & 1 &        &   &   \\
    1 & 0 &        &   &   \\
      &   & \ddots &   &   \\
      &   &        & 0 & 1 \\
      &   &        & 1 & 0 \\
  \end{bmatrix}
  .
\end{equation}
Two stabilizer tableaux are \emph{equivalent} if they represent the same stabilizer group $\mathcal{S}$. One can permute the rows of the matrix without changing the stabilizer group. Also, one can add one row to another while correctly updating signs according to Equation~\eqref{eq:product_of_observables}, which corresponds to multiplying one generator by another. The matrix $\mathsf{U}$ can be degenerate, deleting zero rows from the matrix preserves equivalence. Using Gaussian elimination, one can make any stabilizer tableau non-degenerate and turn it to a row-reduced echelon form (RREF) \cite{Audenaert_2005}.

\subsection{Clifford channels}

A \emph{stabilizer state} is a $n$-qubit quantum state $\rho$ for which there exists a stabilizer group $\mathcal{S}$ such that
\begin{equation}
  \rho = \frac{1}{2^n} \sum_{P\in\mathcal{S}} P.
\end{equation}
The state $\rho$ is pure if and only if $\mathcal{S}$ is maximal, that is it has rank $n$. We will work with both pure and mixed stabilizer states \cite{Aaronson_2004}. Stabilizer states are convenient because it suffices to work with tableaux of their stabilizer groups. The two-qubit Bell state $\frac{\ket{00} + \ket{11}}{\sqrt{2}}$ is stabilizer, and maximally entangled state $\Omega_{AA}$ on two copies of $n$-qubit system $A$ can be written as
\begin{equation}
  \Omega_{AA} = \frac{1}{2^{2\abs{A}}} \sum_{u\in \Int_2^{2n}} P(u)^T\otimes P(u).
\end{equation}
We will call a channel $\Phi : \TrClass(\Hil_A) \to \TrClass(\Hil_B)$ the \emph{Clifford channel} if it maps stabilizer states to stabilizer states. As shown in \cite{Yashin_2025}, Clifford channels can be equivalently characterized as channels with stabilizer Choi states; or as the channels realized by some composition of identity channels, full dephasing channels, stabilizer state preparations, qubit discardings. We argue that Clifford channels describe arbitrary stabilizer operations without classical control \cite{Kliuchnikov_2023, Yashin_2025}. Classical bits can be understood as fully dephased quantum bits, in this setting Clifford channels describe computations on affine Boolean circuits.

Suppose $\Phi : \TrClass(\Hil_A) \to \TrClass(\Hil_B)$ is some Clifford channel and $\sigma_{AB}$ is its Choi state. Since $\sigma$ is a stabilizer state, there is a stabilizer group $\mathcal{S}$ such that
\begin{equation}
  \sigma = \frac{1}{2^{\abs{A}+\abs{B}}} \sum_{P\in\mathcal{S}} \! P \; = \; \frac{1}{2^{\abs{A}+\abs{B}}} \sum_{P\in\mathcal{S}} \! P_A\otimes P_B,
\end{equation}
where $P = P_A\otimes P_B$ is a presentation of Pauli observable $P$ on system $AB$ as a tensor product of Pauli observables on subsystems $A$ and $B$ (there is some freedom in choice of signs). To restore the channel $\Phi$ from its Choi state $\sigma$ one can use the formula
\begin{equation}
  \Phi_{A\to B}[\rho_A] = 2^{\abs{A}} \Tr_{A}\left[\sigma_{AB}\cdot \rho_{A}^T\!\otimes\!I_B \right] = \frac{1}{2^{\abs{B}}} \sum_{P\in\mathcal{S}} \Tr\left[ \rho_A P_{A}^T \right] P_B.
\end{equation}
This equation shows how the stabilizer group $\mathcal{S}$ of Choi state defines the action of the channel $\Phi$. As discussed above, $\mathcal{S}$ can be stored as a bit matrix $[\mathsf{U}|\mathsf{c}]$, so also all the information about the channel $\Phi$ is stored in this matrix.

For the sake of mathematical beauty we will instead store this information in slightly modified form. Because the input system $A$ and output system $B$ are distinguished, it is more natural to represent the channel $\Phi$ as a $k\!\times\!(2\abs{A}+2\abs{B}+1)$ bit matrix $[\mathsf{U}_A|\mathsf{U}_B|\mathsf{c}]$, separating between the columns correponding to $A$ and $B$. The modification is that we now interpret a bit string as a Pauli superoperator insted of Pauli operator: row $[u_A|u_B|c]$ corresponds to a superoperator $\Pi(u_A|u_B|c) : \TrClass(\Hil_A) \to \TrClass(\Hil_B)$ defined as
\begin{equation}
  \Pi(u_A|u_B|c)[\rho_A] = (-1)^c \, 2^{\abs{A}}\Tr[\rho_A\, P(u_A)] P(u_B)
\end{equation}
for $u_A \in \Int_2^{2\abs{A}}, u_B\in\Int_2^{2\abs{B}}, c\in\Int_2$. The Choi operator of superoperator $\Pi(u_A|u_B|c)$ is
\begin{equation}
  \Id_{A}\otimes \Pi(u_A | u_B | c) [\Omega_{AA}] = P^{T_A}(u_A,u_B|c)_{AB} = P(u_A,u_B|c\oplus\innerp{z_A}{x_A})_{AB},
\end{equation}
where the superindex ${}^{T_A}$ means partial transposition over system $A$.
\begin{definition}
By a \emph{stabilizer tableau of a Clifford channel $\Phi_{A \to B}$} we mean a Boolean matrix $[\mathsf{U}_A|\mathsf{U}_B|\mathsf{c}]$ that encodes a set of superoperators $\mathcal{S}$ such that
\begin{equation}
  \Phi[\rho] = \frac{1}{2^{\abs{A}+\abs{B}}} \sum_{\Pi(u_A|u_B|c) \in \mathcal{S}} \Pi(u_A|u_B|c)[\rho]
  = \frac{1}{2^{\abs{B}}} \sum_{\Pi(u_A|u_B|c) \in \mathcal{S}} (-1)^c \Tr[\rho\, P(u_A)] P(u_B).
\end{equation}
\end{definition}

As in the case of stabilizer states, equivalent tableaux can define the same Clifford channel, so let us study how they behave under row operations. Obviously, one freely can swap rows or delete zero rows. In order to understand how to sum rows in a tableau $[\mathsf{U}_A|\mathsf{U}_B|\mathsf{c}]$, let us compute the product of two Choi operators:
\begin{equation}
  P^{T_A}(u_A,u_B,c) \cdot P^{T_A}(u_A',u_B',c') = i^{\beta_A(u_A',u_A)+\beta_B(u_B,u_B')} P^{T_A}(u_A\oplus u_A',u_B\oplus u_B',c\oplus c').
\end{equation}
That means, if we want to make some row updates on $[\mathsf{U}_A|\mathsf{U}_B|\mathsf{c}]$, we can add rows in sign-free part $[\mathsf{U}_A|\mathsf{U}_B]$ as usual, but the sign update will depend on the separation between input system $A$ and output system $B$:
\begin{equation}
  \text{adding row } [u_A|u_B|c] \text{ to } [u_A'|u_B'|c'] \text{ results in } [u_A\oplus u_A'|u_B\oplus u_B'|\,c\oplus c' \oplus {\scriptstyle \frac{\beta(u_B,u_B')-\beta(u_A,u_A')}{2} } ].
\end{equation}
Row summation defines a multiplication between Pauli superoperators $\Pi$, we wonder if there is any natural interpretation for it except of ``the multiplication of Choi operators''.

Given a Boolean matrix $[\mathsf{U}_A|\mathsf{U}_B|c]$, when does it correspond to some Clifford channel? Firstly, the matrix should be feasible in the sense that one cannot obtain row $[0|0|1]$ as a result of row operations. (Or rather, infeasibility implies that the channel is trivial.) Secondly, the sign-free part $[\mathsf{U}_A|\mathsf{U}_B]$ should be symplectic. Thirdly, the trace preserving condition $\Tr\circ\Phi = \Tr$ implies that columns of $\mathsf{U}_A$ linearly depend on columns of $\mathsf{U}_B$, so one can always make the number of rows to be $k \leq 2\abs{B}$. These conditions are also sufficient, because they define correct tableau for stabilizer Choi state.

Suppose two Clifford channels $\Phi_{A\to B}$ and $\Phi'_{A'\to B'}$ are represented by stabilizer tableaux $[\mathsf{U}_A|\mathsf{U}_B|\mathsf{c}]$ and $[\mathsf{U}'_{A'}|\mathsf{U}'_{B'}|\mathsf{c}']$ respectively. Then their tensor product $\Phi\otimes \Phi'$ can be represented by the direct sum of two tableaux:
\begin{equation}
  \left[
  \begin{array}{cc|cc|c}
    \mathsf{U}_A & 0                & \mathsf{U}_B & 0                & \mathsf{c} \\
    0            & \mathsf{U}'_{A'} & 0            & \mathsf{U}'_{B'} & \mathsf{c}' \\
  \end{array}
  \right].
\end{equation}

Let us discuss how stabilizer tableaux are composed. The composition of two Pauli superoperators is
\begin{equation}
  \Pi(u'_B|u'_C|c') \circ \Pi(u_A|u_B|c) = 2^{2\abs{B}} \delta_{u_B,u'_B}\Pi(u_A|u'_C|c\oplus c'),
\end{equation}
where $\delta_{u_B,u'_B}$ is the Kronecker delta-symbol. Given two Clifford channels $\Phi_{A\to B}$ and $\Phi'_{B\to C}$ represented as sums
\begin{equation}
  \Phi_{A\to B} = \frac{1}{2^{\abs{A}+\abs{B}}} \sum_{(u_A|u_B|c)} \Pi(u_A|u_B|c), \qquad
  \Phi'_{B\to C} = \frac{1}{2^{\abs{B}+\abs{C}}} \sum_{(u'_B|u'_C|c')} \Pi(u'_B|u'_C|c'),
\end{equation}
their composition is
\begin{equation}
  (\Phi' \circ \Phi)_{A\to C} = \frac{1}{2^{\abs{A}+\abs{C}}} \sum_{\substack{(u_A|u_B|c) \\ (u'_B|u'_C|c') \\ u_B = u'_B}} \Pi(u_A|u'_C|c\oplus c').
\end{equation}
We see that the tableau of the composition $\Phi'\circ \Phi$ is generated by all bit strings $[u_A|v_C|c\oplus s]$, where $[u_A|u_B|c]$ and $[u'_B|u'_C|c]$ are bit strings from both stabilizer tableaux such that $u_B = v_B$. Finding a basis for such strings (a basis in the intersection of two vector spaces) is a problem of Boolean linear algebra \cite{Damm_1990}, we further discuss this problem below.

\section{Explaining stabilizer circuit simulation} \label{sec:explaining}

In this Section we apply the introduced framework to explain and analyse stabilizer circuits simulation. First, we discuss stabilizer tableau descriptions of commonly used operations and the costs of manipulating them. Then, we consider rewriting a given stabilizer circuit as a diagram of stabilizer tableaux and the problem of contracting this diagram. Some contraction strategies reproduce known algorithms for stabilizer simulation. Finally, we summarise the ideas and give an outlook.

\subsection{Elementary operations}

Quantum circuits are assembled from a set of elementary operations -- state preparations, gates, measurements. Common stabilizer operations with corresponding tableaux are listed in Table~\ref{tab:elementary_operations}. Note that we ignore working with classical wires separately from quantum wires. As mentioned earlier, it is convenient for us to consider classical bits as dephased quantum bits. That means, we consider a classical wire as a quantum wire without any coherence, in terms of tableau it means that the $X$-column corresponding to this qubit is always zero. Under this convention the destructive $Z$-measurement of a qubit is exactly $Z$-dephasing. If needed, one can consider classical wires to represent special type of columns. This fine tuning can potentially save some space when storing the tableau, but it would complicate our discussion.

\begin{table}[p]
  \caption{The list of basic stabilizer operations from input system $A$ to output system $B$ often used to assemble stabilizer circuits, their common circuit depiction and corresponding stabilizer tableaux. Empty columns or rows are indicated by dot $[\cdot]$.}
  \label{tab:elementary_operations}
  \begin{ruledtabular}
  \begin{tabular}{ccccc}
    Clifford channel $\Phi_{A\to B}$ & $\abs{A}$ & $\abs{B}$ & Circuit depiction & Stabilizer tableau $[\mathsf{U}_A|\mathsf{U}_B|\mathsf{c}]$ \\[0.3em]
    \hline\\[-0.7em]
    chaotic state $\chi$ preparation & 0 & 1 &
    \begin{quantikz} \lstick{$\chi$} & \end{quantikz} &
    $\left[ \cdot \right]$ \\[1.2em]
    initial state $\ket{0}$ preparation & 0 & 1 &
    \begin{quantikz} \lstick{$\ket{0}$} & \end{quantikz} &
    $\left[ \begin{array}{c|cc|c} \cdot & 1 & 0 & 0\end{array} \right]$ \\[1.2em]
    state $\ket{1}$ preparation & 0 & 1 &
    \begin{quantikz} \lstick{$\ket{1}$} & \end{quantikz} &
    $\left[ \begin{array}{c|cc|c} \cdot & 1 & 0 & 1\end{array} \right]$ \\[1.2em]
    state $\ket{+}$ preparation & 0 & 1 &
    \begin{quantikz} \lstick{$\ket{+}$} & \end{quantikz} &
    $\left[ \begin{array}{c|cc|c} \cdot & 0 & 1 & 0\end{array} \right]$ \\[1.2em]
    state $\ket{-}$ preparation & 0 & 1 &
    \begin{quantikz} \lstick{$\ket{-}$} & \end{quantikz} &
    $\left[ \begin{array}{c|cc|c} \cdot & 0 & 1 & 1\end{array} \right]$ \\[1.2em]
    qubit discarding $\Tr$ & 1 & 0 &
    \begin{quantikz} &\ground{} \end{quantikz} &
    $\left[ \cdot \right]$ \\[1.2em]
    identity channel $\Id$ & 1 & 1 &
    \begin{quantikz} & & \end{quantikz} &
    $\left[ \begin{array}{cc|cc|c} 1 & 0 & 1 & 0 & 0 \\ 0 & 1 & 0 & 1 & 0\end{array} \right]$ \\[1.2em]
    $Z$-dephasing channel $\Deph_Z$      & 1 & 1 &
    \begin{quantikz} &\gate{\Deph_Z} & \end{quantikz} &
    $\left[ \begin{array}{cc|cc|c} 1 & 0 & 1 & 0 & 0 \end{array} \right]$ \\[1.2em]
    $X$-dephasing channel $\Deph_X$      & 1 & 1 &
    \begin{quantikz} &\gate{\Deph_X} & \end{quantikz} &
    $\left[ \begin{array}{cc|cc|c} 0 & 1 & 0 & 1 & 0 \end{array} \right]$ \\[1.5em]
    Pauli gate $Z$ & 1 & 1 &
    \begin{quantikz} &\gate{Z} & \end{quantikz} &
    $\left[ \begin{array}{cc|cc|c} 1 & 0 & 1 & 0 & 0 \\ 0 & 1 & 0 & 1 & 1\end{array} \right]$ \\[1.5em]
    Pauli gate $X$ & 1 & 1 &
    \begin{quantikz} &\gate{X} & \end{quantikz} &
    $\left[ \begin{array}{cc|cc|c} 1 & 0 & 1 & 0 & 1 \\ 0 & 1 & 0 & 1 & 0\end{array} \right]$ \\[1.5em]
    Hadamard gate $H$ & 1 & 1 &
    \begin{quantikz} &\gate{H} & \end{quantikz} &
    $\left[ \begin{array}{cc|cc|c} 1 & 0 & 0 & 1 & 0 \\ 0 & 1 & 1 & 0 & 0\end{array} \right]$ \\[1.5em]
    phase gate $S$ & 1 & 1 &
    \begin{quantikz} &\gate{S} & \end{quantikz} &
    $\left[ \begin{array}{cc|cc|c} 1 & 0 & 1 & 0 & 0 \\ 0 & 1 & 1 & 1 & 0\end{array} \right]$ \\[1.7em]
    gate $\CNOT$ & 2 & 2 &
    \begin{quantikz} &\ctrl{1} & \\ & \targ{} & \end{quantikz} &
    $\left[ \begin{array}{cccc|cccc|c}
      1 & 0 & 0 & 0 & 1 & 0 & 0 & 0 & 0\\
      0 & 1 & 0 & 0 & 0 & 1 & 0 & 1 & 0\\
      0 & 0 & 1 & 0 & 1 & 0 & 1 & 0 & 0\\
      0 & 0 & 0 & 1 & 0 & 0 & 0 & 1 & 0\\
    \end{array} \right]$ \\[2.5em]
    gate $CZ$ & 2 & 2 &
    \begin{quantikz} &\ctrl{1} & \\ & \ctrl{0} & \end{quantikz} &
    $\left[ \begin{array}{cccc|cccc|c}
      1 & 0 & 0 & 0 & 1 & 0 & 0 & 0 & 0\\
      0 & 1 & 0 & 0 & 0 & 1 & 1 & 0 & 0\\
      0 & 0 & 1 & 0 & 0 & 0 & 1 & 0 & 0\\
      0 & 0 & 0 & 1 & 0 & 0 & 1 & 1 & 0\\
    \end{array} \right]$ \\[1.2em]
  \end{tabular}
  \end{ruledtabular}
\end{table}

Let us discuss the costs of storing and working with stabilizer tableaux. Suppose $\Phi_{A\to B}$ is a stabilizer operation with stabilizer tableau $[\mathsf{U}_A|\mathsf{U}_B|\mathsf{c}]$ of size $k\!\times\!(2n+1)$ where $n = \abs{A}+\abs{B}$. Any tableau will have no more than $\leq 2\abs{B}$ linearly independent rows. One requires $\BigO(k n)$ bits of memory to store the tableau, or less in case the tableau is sparse. One can do row operations over the tableau: swapping two rows requires $\BigO(1)$ time; adding one row to another (together with computing non-linear sign update) requires $\BigO(n)$ operations and can be parallelized. Any tableau can be taken to a row-reduced echelon form (RREF) \cite{Audenaert_2005}. It will often be sufficient to reduce only a single row in a tableau, or reduce either $A$ part or $B$ part. When using the procedure of Gaussian elimination, one requires $\BigO(k)$ row operations to reduce a single column, or $\BigO(k^2)$ row operations to compute full RREF in $\BigO(k^2 n)$ time. For enormously huge matrices one might consider using advanced fast multiplication algorithms instead of Gaussian elimination \cite{Strassen_1969, ONeil_1973, Albrecht_2010, Albrecht_2011}, which propose asymptotical advantages but are impractical for small-scaled matrices.

Taking tensor product of Clifford channels corresponds to taking direct sum of their tableaux, which does not require additional resources. To obtain the tableau of the composition of channels $\Phi_{A\to B}$ with $\Phi'_{B\to C}$, one can compute RREF on system $B$ for both tableaux and find their intersection, in the worst case this will take no more than $\BigO((\abs{A}+\abs{B}+\abs{C})^3)$ time, but often is much more
efficient. Let us discuss some cases.
\begin{itemize}
  \item The operation of discarding a qubit $\Tr$ has no rows in it's tableau, so discarding all the qubits takes $\BigO(1)$ time. Suppose we want to discard a single qubit from the set of qubits, meaning to compose $\Phi_{A\to B}$ with $\Tr^1\otimes\Id^{\abs{B}-1}$. One should perform Gaussian elimination over the column of the discarded qubit in $\mathsf{U}_B$, then delete the rows that are non-zero on this qubit, the procedure takes $\BigO(k n)$ time. In contrast, discarding all but one qubit takes $\BigO(k^2 n)$ time.
  \item A unitary Pauli operation $\rho \mapsto P(u) \rho P(u)^\dag$ has stabilizer tableau $[\,\mathsf{I}\,|\,\mathsf{I}\,|\,\mathsf{J} u^T]$, where $\mathsf{I}$ is the identity matrix and $\mathsf{J}$ is the matrix of symplectic form. Thus, to compose $\Phi$ with Pauli operation it suffices to compute $\mathsf{U}_B \mathsf{J}\,u^T$ (or $\mathsf{U}_A \mathsf{J}\,u^T$ for pre-composition), which might take $\BigO(k n)$ time. If the Pauli gate is one-qubit, it takes $\BigO(k)$ time or less.
  \item Suppose $U$ is a Clifford unitary. The stabilizer tableau of $\rho \mapsto U \rho U^\dag \leadsto$ can be represented as $[\mathsf{I}|\mathsf{S}|\mathsf{c}]$, where $\mathsf{I}$ is the identity matrix and $\mathsf{S}$ is some symplectic matrix $\mathsf{S}\,\mathsf{J}\,\mathsf{S}^T = \mathsf{J}$. One can efficiently invert the Clifford unitary in time $\BigO(n^2)$ using $\mathsf{S}^{-1} = \mathsf{J}\mathsf{S}\mathsf{J}$, so the RREF over output subsystem gives $[\mathsf{S}^{-1}|\mathsf{I}|\mathsf{S}^{-1}\mathsf{c}]$. Relatedly, the tableau of $\rho \mapsto U^\dag \rho U$ is $[\mathsf{S}|\mathsf{I}|\mathsf{c}]$. Thus, one can efficienlty compose any stabilizer operation with global Clifford unitary in time $\BigO(k n^2)$ using matrix multiplication, while applying one- or two-qubit unitaries require $\BigO(k)$ time.
  \item The channel of measuring Pauli observable $P(u|c)$ taking $\abs{B}$ qubits and producing $1$-bit outcome has stabilizer tableau $[u|1\,0|c]$. To find the outcome of the measurement (it can be either deterministic or unifromly random), one should check if $u$ lies in the linear span of rows $\mathsf{U}_B$ or not. This can be done by taking $\mathsf{U}_B$ to RREF in $\BigO(k^2 n)$ time.
\end{itemize}

Let us briedly touch on the procedure of post-selection. We can include it in the theory by allowing the columns of input system $A$ not to depend on columns of output system $B$. For example, the trace-decreasing map $\rho \mapsto \Tr[\rho\,\frac{I+P(u|c)}{2}]$, the meaning of which is to post-select a state on observable outcome $P(u|c) = 1$, is expressed as a tableau $[u|\cdot|c]$. Using such tableaux can be useful in the problem of sampling from the measurement result (weak simulation), but can also result in infeasibilities $[0|0|1]$ which imply that the channel is trivial.

\subsection{Global simulation}

Suppose we are given some non-adaptive stabilizer circuit $\mathcal{C}$, that is a network of composed elementary stabilizer operations. More formally, the circuit can be seen as a directed acyclic graph with labeled vertices, or as a diagram in the category of multiqubit systems and stabilizer operations between them. Let us rewrite each element of a circuit as a stabilizer tableau. That is, let us transform the circuit $\mathcal{C}$ to a directed acyclic graph with stabilizer tableau at each vertex, the directed edges being the qubit wires from the output of one channel to the input of another. There is a freedom in updating the resulting diagram using the following rewriting rules:
\begin{enumerate}
  \item One can make row operations on tableaux without changing the structure of a graph.
  \item One can take a direct sum of two parallel stabilizer tableaux, glueing together two vertices.
  \item One can compose two stabilizer operations, providing a contraction over some edge.
\end{enumerate}
Such rewriting rules do not change the overall channel that this circuit describes. Using these rules, one can contract the circuit $\mathcal{C}$ to one vertex, giving a final stabilizer tableau for the whole circuit.

Let us consider the case when the circuit $\mathcal{C}$ has no input wires and that all output is classical (there are measurements at the end of each wire). After constructing the diagram and contracting it to a single vertex, we get a stabilizer tableau of form $[\cdot|\mathsf{M}|\mathsf{c}]$, where $\mathsf{M}$ is a matrix with zero $X$-columns, suppose we computed it's rank $r$. Then, the probability of an $n$-bit outcome $x$ to occur equals $p(x) = 2^{n-r}$ if $\mathsf{M}x = \mathsf{c}$ (here we ignore $X$-bits) and equals $p(x)=0$ otherwise. The problem of computing the probability $p(x)$ of a given bit outcome $x$ is called \emph{strong simulation}. The problem of strong simulation of non-adaptive stabilizer circuits is known to be complete for the class of problems solvable on affine Boolean circuits \cite{Aaronson_2004}, and is equivalent to the usual linear-algebraic problems such as solving the system of Boolean linear equations. And indeed, each contraction step of the algorithm is Gaussian elimination.

There may be many efficient strategies to contract the diagram, creating a room for optimization of simulation algorithms. In fact, the correct way to contract diagrams is an important research problem in the fields of tensor networks \cite{Biamonte_2017, Montangero_2018} and diagrammatic calculi \cite{Wetering_2020}. Let us discuss two classical strategies for such contraction. The first one is the most straightforward, introduced by Gottesman and Knill \cite{Gottesman_1997,Nielsen_2010}. It consists of working only with stabilizer tableau of a state and updating it online with gates and measurements. This is equivalent to a strategy of choosing an $n$-qubit pure initial state and contracting a diagram to it step-by-step. Updating by local unitary gate requires $\BigO(n)$ time, measuring an observable requires Gaussian elimination and $\BigO(n^3)$ time. A more involved approach due to Aaronson and Gottesman \cite{Aaronson_2004} is to store the information about the state in \emph{extended stabilizer tableau} involving the stabilizers and destabilizers. In our framework, the idea is equivalent to representing a stabilizer state $\ket{\psi}$ as a pair $\ket{\psi} = U\ket{0}$ and to store the tableau of a Clifford unitary channel $\rho \mapsto U\rho U^\dag$. Contracting $U$ with local gates requires $\BigO(n)$ time, but finding the $1$-bit measurement outcome is faster because taking $U$ into RREF form on output takes only $\BigO(n^2)$ time.

To do weak simulaiton (sampling from the outcomes), one can sequentially do the following steps: compute the result of a measurement, take a sample from it and post-select on the outcome. To sample from concrete measurement, one should contract full diagram while forgetting all other outcomes. The tricky part is choosing the right strategy to sequentially simplify the diagram. Strategies of Gottesman and Aaronson-Gottesman for weak simulation essentially follow the described procedure.

\subsection{Outlook}

To sum up, in this note we have developed a stabilizer tableau formalism for arbitrary qubit stabilizer operations and have noticed that a stabilizer circuit can be rewritten as a diagram of stabilizer tableaux. Contracting an edge in the diagram is a problem of Boolean linear algebra and can be reduced to Gaussian elimination. To simulate a circuit means to contract the diagram, so stabilizer circuit simulation is essentially a sequence of Gaussian eliminations. There may be various strategies to contract a circuit. Let us now comment on what improvements could be done.

First of all, one might consider trying to check our claims in practice, that is to write a program that works with stabilizer tableaux of Clifford channels, that represents a circuit as a diagram of tableaux and that contracts this diagram using some strategy. This program could be compared with existing pacakages \cite{Aaronson_2004,Gidney_2021}, we predict the performance will be similar. It might be interesting to test the performance on random stabilizer circuits with measurements and noises. As discussed above, the problem of weak simulation can be reduced to strong simulation bit-by-bit. At the same time, there exist other algorithms for sampling, such as gate-by-gate sampling \cite{Bravyi_2022}. We expect that our framework allows for comparing the performance of sampling procedures, and that it may be useful for studying error propagation in stabilizer circuits.

There are works trying to create correct stabilizer tensor networks theory \cite{Masot_Llima_2024, Nakhl_2025, Backens_2017, Kissinger_2022, Cao_2022}. Note that our framework also gives an example of such theory, it might be useful to compare our approach with existing ones. One drawback of our framework is that it is phase-independent, so it cannot include simulation of close-to-stabilizer magic operations by stabilizer decompositions \cite{Bravyi_2016,Bravyi_2016_2,Bravyi_2019}.

Finally, it is rather straightforward to generalize the framework from qubit theory to qudit theory \cite{Gottesman_1999, Beaudrap_2013, Gheorghiu_2014}. We suspect this will actually work for any stabilizer theory over finite abelian groups, we plan to investigate it in the future \cite{Yashin_TBA}.

\section*{Acknowledgements}
This work was performed at the Steklov International Mathematical Center and supported by the Ministry of Science and Higher Education of the Russian Federation (agreement no. 075-15-2025-303).

\nocite{Kay_2023} 
\bibliography{./bibliography.bib}

\end{document}